\documentclass[11pt,epsfig]{article}
\usepackage{epsf,amssymb}

\newcommand\hb{\hbar}
\begin{document}
\begin{titlepage}
\begin{flushright}
DFTT 13/99\\
March 1999
\end{flushright}
\vskip0.5cm
\begin{center}
{\Large\bf
The confining string and its breaking in QCD
}\\
\end{center}
\vskip 0.6cm
\centerline{F. Gliozzi$^{a}$ and P. Provero$^{b,a}$}
\vskip 0.6cm
\centerline{\sl $^{a}$ Dipartimento di Fisica
Teorica dell'Universit\`a di Torino and}
\centerline{\sl Istituto
Nazionale di Fisica Nucleare, Sezione di Torino}
\centerline{\sl via P.Giuria 1, I--10125 Torino, Italy}
\vskip0.6cm
\centerline{\sl $^{b}$ Dipartimento di Scienze e Tecnologie Avanzate}
\centerline{\sl Universit\`a del Piemonte Orientale, Alessandria, Italy
\footnote{e--mail:gliozzi, provero@to.infn.it}}
\vskip 0.6cm
\begin{abstract}
We point out that the world sheet swept by the confining
string in presence of dynamical quarks can belong to two different
phases, depending on the number of charge species and the
quark masses. When it lies in the normal phase (as opposed to
the tearing one) the string breaking is invisible in the
Wilson loop, while is manifest in operators composed of
disjoint sources, as observed in many  numerical experiments.
We work out an explicit formula for the correlator of Polyakov
loops at finite temperature,  which is then  compared with
recent lattice data, both in the quenched case and in
presence of dynamical quarks. The analysis in the quenched
case shows that the free bosonic string model
describes accurately the data for distances larger
than $\sim 0.75$ fm. In the unquenched case we derive
predictions on the dependence of the static potential on
the temperature which are compatible with the lattice data.

\end{abstract}
\vskip .5cm
\hrule
\end{titlepage}
\setcounter{footnote}{0}
\def\thefootnote{\arabic{footnote}}
\section{Introduction}
The  confining interaction of a pair of static sources
in non Abelian gauge theories
is mediated by a thin flux tube, or string,
joining the two sources.
At large distance, the energy of the flux tube is
proportional to its length,
hence the static potential grows linearly.
\par
When matter in the fundamental representation is added
to the system, one expects that this potential flattens
at some distance $r_o$, where the string breaks to form
pairs of matter particles which screen the confining
potential. The broken string state describes a bound state
of a static color source (the fixed end of the string)
and a dynamical matter field (the free end of the string).
A similar screening is expected even in pure Yang- Mills
theory for adjoint sources, since in this case
the adjoint string may terminate on  dynamical gluons.

Despite many efforts, these string breaking effects
have proved elusive  in standard analyses
of large Wilson loops, both in QCD with dynamical
fermions \cite{data}
and in pure Yang-Mills theory with adjoint
sources in 3+1 dimensions \cite{CM} and in 2+1 dimensions
\cite{PT}.

On the contrary, clear signals of string breaking
 have been observed in studies
where the basis of operators has been enlarged in
order to find a better overlap to the true
ground state, following a method originally advocated in Ref.~\cite {gmr}. 
In this way, fundamental string breaking has
been found in SU(2) Higgs model in
2+1 \cite{PW} and 3+1 \cite{KS} dimensions, and
adjoint string breaking in 2+1 SU(2) pure gauge theory
\cite{Sa,PWa}.

The outcome of these analyses is twofold.
On one hand  the Wilson loop
appears to have in general very poor overlap on
the broken string state, the only exception  being
 observed up to now in the 2+1 dimensional SU(2)
gauge theory with two flavors of staggered fermions
\cite{tr}. On the other hand it turns out that the
operators with a good overlap with the broken
string state have as a common feature
the presence of two disjoint source lines.
This is  not only  true in the above mentioned cases
at zero temperature,
but it is also evident  in the recent observation of
string breaking at finite
temperature QCD with dynamical fermions, where
the static potential is extracted from the
correlator of two disjoint Polyakov loops \cite{detar}.

In this work we will show that in the usual string
description of confinement \cite{genstring}
one can find a simple explanation of such a
relationship between the overlap properties
and the number of disjoint static sources.
In particular we shall derive an asymptotic functional
form of the unquenched finite temperature potential
which is then successfully compared with the QCD data
of Ref.~\cite{detar}.

%The organization of the paper is as follows.
%Our starting point
%In section \ref{topo}, using the results of a
A solvable prototype of string breaking \cite{kaza}
suggests that the
world sheet swept by the string in its time evolution
can exist in two different phases, known as tearing
and normal phases. The question
of the existence of a finite overlap of the Wilson
operator with  the broken string state can  be
reformulated (see section \ref{topo}) as the problem
of finding the phase of the string world sheet.
Indeed in the former phase the world sheet is  torn
by large holes corresponding to pair
creation of charged matter, hence  the Wilson loop is
expected to fulfil the perimeter law at large distances.
On the contrary, in the normal phase the string breaking
is balanced by the inverse process of string soldering,
so that adding  dynamical matter to
the system yields  simply a renormalization of the
string tension. In particular the Wilson loop behaves
exactly like in the quenched model and no macroscopic
string breaking is visible.
\par
However, a simple topological argument shows 
that the situation drastically changes when one considers, 
instead of the Wilson loop, operators constructed out of 
{\em disjoint} static sources.
Besides the usual contribution to the confining
potential there is a new term which is necessarily
absent in the quenched case and which survives at large
source separations.
This term makes string breaking effects easily detected on the lattice.
\par
The remaining sections are organized as follows.
In section \ref{string} we review some known
bosonic string formulae about the asymptotic infrared
behaviour of gauge operators and derive the new
term contributing to correlator of two Polyakov
loops in the unquenched case. Since there are no
recent studies on the string  effects
in the quenched approximation of QCD, we devote 
section \ref{quen} to an analysis of the
quenched data of Ref.~\cite{detar}. It turns out that
the bosonic string accurately describes the lattice
data for distances larger than $\sim 0.75$ fm. In section
\ref{unqu} the analysis is extended to the unquenched case
where we check that string predictions on the temperature
dependence of the potential are compatible with the
lattice data of Ref.~\cite{detar}. Finally in section
\ref{conc} we draw some concluding remarks.

\section{A topological argument}
\label{topo}
Let us start by considering
the standard Wilson loop. In absence of charged dynamical fields
this loop acts as a fixed boundary of the surface associated to the
world sheet swept by the confining string. When matter fields are
added, any number of holes of any size may appear on this surface,
reflecting the pair creation of dynamical charged particles.
If these are light, the holes behave like free boundaries.
The total string contribution to the Wilson operator is
then the sum over all possible insertions of multiconnected
boundaries of the world sheet. It can be written
diagrammatically as a loop expansion (see Fig.~1).

There are two quantities
controlling the number and the size of these holes. One is
the mass of the matter fields (it is easier to create
lighter particles). The other is the number of charge species.
For a theory with $N_f$ flavors and
$N_c$ colors there is a factor of $N_fN_c$ for each hole. Likewise
in the adjoint string every hole is accompanied by a factor of
$N_c^2-1$.

For light particles and $N_fN_c$ large enough, we expect
that the string world sheet is dominated by
configurations with a large number of relatively small
holes compared with the size of the Wilson loop.

These holes do not influence the area law of the Wilson loop
and their effect can be absorbed into the renormalization
of the string tension, according to a string mechanism originally
proposed in Ref.~\cite{ademo}. Under the circumstances there
is no way to observe the string breaking
 through the study of large Wilson loops. This may explain
 why this phenomenon has been so elusive both for fundamental
 and adjoint string in any dimension.
{\begin{figure}[ht]
\[\begin{array}{l}
\epsfxsize=.57\linewidth\epsfbox{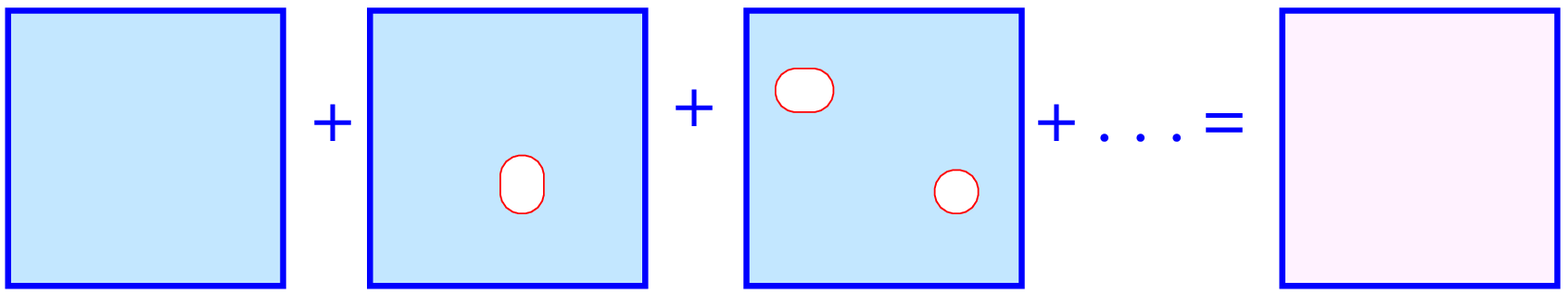}
%{}&
%\epsfxsize=.37\linewidth\epsfbox{5-dual.eps}
\\
\parbox[t]{.57\linewidth}{\small Figure 1. Loop expansion
of the rectangle.}
%{}~&~
%\parbox[t]{.47\linewidth}{\small 1b) $dual ~polytope$.}
\end{array}\]
\end{figure}}
 However it is worth noting, as mentioned in the introduction,
 that there is a solvable matrix model \cite {kaza} describing
 idealized random surfaces with dynamical holes
 (or open strings embedded  in zero dimension), where
 besides the above described "normal" phase dominated by holes of small
 size there is also another phase, characterized by a spontaneous
 tearing of the surface, due to the formation of large
 holes growing with the size of the surface (see Fig.~2);
 as a consequence we expect that the
 associated Wilson loop should fulfil the
perimeter law.\footnote{There is also a third phase separating the
above two, where both the world sheet (the glue) and the holes
(the charged matter) are large and in competition. This phase could play
a role at the deconfining temperature.}
{\begin{figure}[ht]
\[\begin{array}{l}
\epsfxsize=.37\linewidth\epsfbox{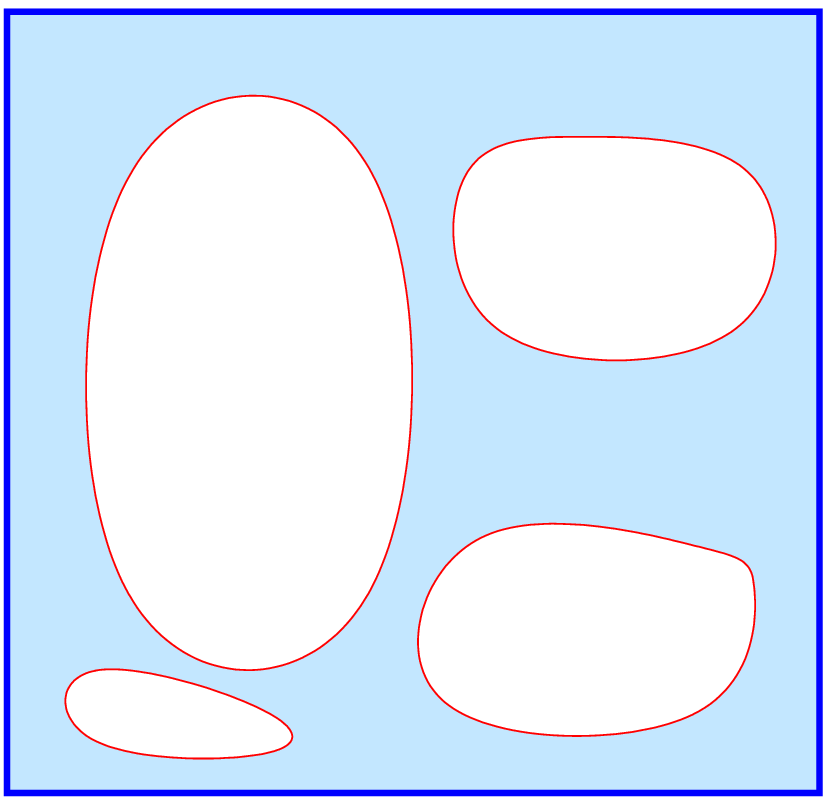}
%{}&
%\epsfxsize=.37\linewidth\epsfbox{5-dual.eps}
\\
\parbox[t]{.50\linewidth}{\small Figure 2. Configuration of a torn
surface}
%{}~&~
%\parbox[t]{.47\linewidth}{\small 1b) $dual ~polytope$.}
\end{array}\]
\end{figure}}

What is the phase of the world sheet of the confining string?
The observed extremely poor overlap of the Wilson loop with
the broken string state seems to suggest that it
belongs to the normal phase\footnote{Of course the numerical data  cannot exclude
the possibility that such a poor overlap is a finite
size effect and that larger Wilson loops will eventually
reveal a spontaneous tearing of the world sheet.},
with a possible exception for
the SU(2) gauge theory in 2+1 dimensions with
dynamical fermions \cite{tr}. Perhaps in this case
the quarks are not light enough.

 For the operators with more than one connected source
 line the loop expansion has a topologically different
 form. Take for instance a pair of Polyakov loops of a
 gauge theory with dynamical fermions at finite
 temperature. From the point of view of the effective string,
 they are the two fixed boundaries of a
 cylinder (see Fig.~3). The loop expansion
 splits into two different series, because starting at two-loop
 level there are configurations in which the two fixed
 boundaries are not connected by the string world sheet.

 If the system is  in the normal phase the sum over
 the loop insertions can be still absorbed into  the
 renormalization of the string tension, but now the loop
 expansion is the sum of two different terms. That
 with only one connected world sheet has no overlap
 to the broken string state but decays exponentially
 with the distance between the fixed boundaries, thus only
 the term with two disjoint  pieces survives, which
 of course is expected to have a large overlap to
 the broken string state.
 {\begin{figure}[ht]
\[\begin{array}{l}
\epsfxsize=.57\linewidth\epsfbox{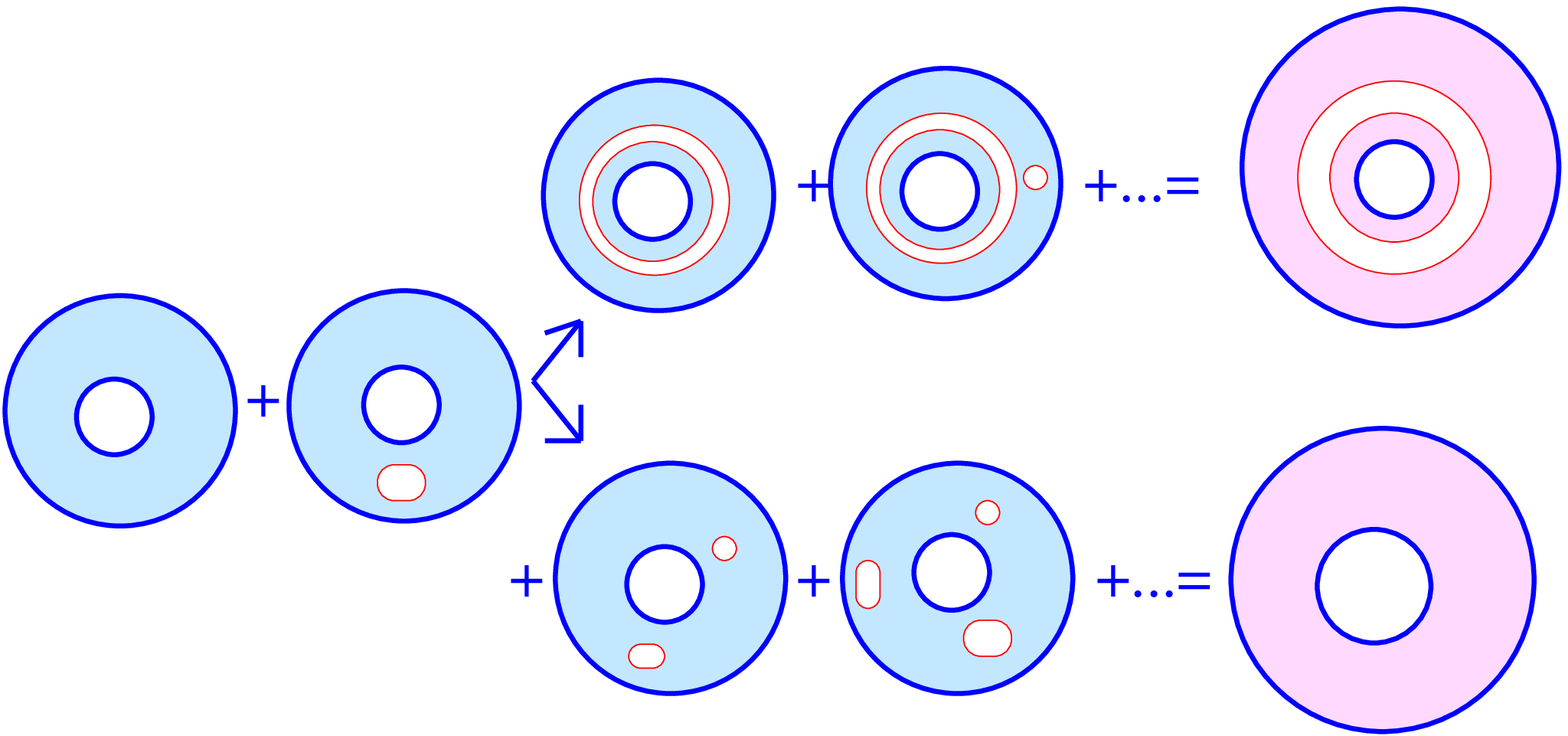}
%{}&
%\epsfxsize=.37\linewidth\epsfbox{5-dual.eps}
\\
\parbox[t]{.57\linewidth}{\small Figure 3. Loop expansion
for the cylinder.}
%{}~&~
%\parbox[t]{.47\linewidth}{\small 1b) $dual ~polytope$.}
\end{array}\]
\end{figure}}
In the case of light matter fields this term can be
written as a square $[Z_{DN}(R_o,L_t)]^2$,
where $Z_{DN}$ denotes the contribution of a cylindric
world sheet with fixed (or Dirichlet) boundary
conditions (b.c.) on the Polyakov line of length $L_t$ and
free (or Neumann) b.c. on the other side (representing a
dynamical charge) placed at a mean distance $R_o$ from the
Polyakov loop. Clearly $R_o$ fixes the scale of the
string breaking. Owing to the simple geometry of this term,
we shall explicitly evaluate its functional form in the next
section.

Similar considerations apply to operators used to see string
breaking at zero temperature. In such a case the fixed
boundary is not necessarily closed because it may terminate
on charged fields. The rule to associate a string
world sheet to these operators is simply to connect the
end points of the fixed boundaries (Wilson lines) with paths
associated to the free end of the string, representing
the world lines of the dynamical particles. For instance,
using the diagrammatic representation of Ref.~\cite{KS},
the map to the string picture of other operators used to
determine the static potential in the SU(2) Higgs model
is drawn in Fig.~4.
 {\begin{figure}[ht]
\[\begin{array}{l}
\epsfxsize=.57\linewidth\epsfbox{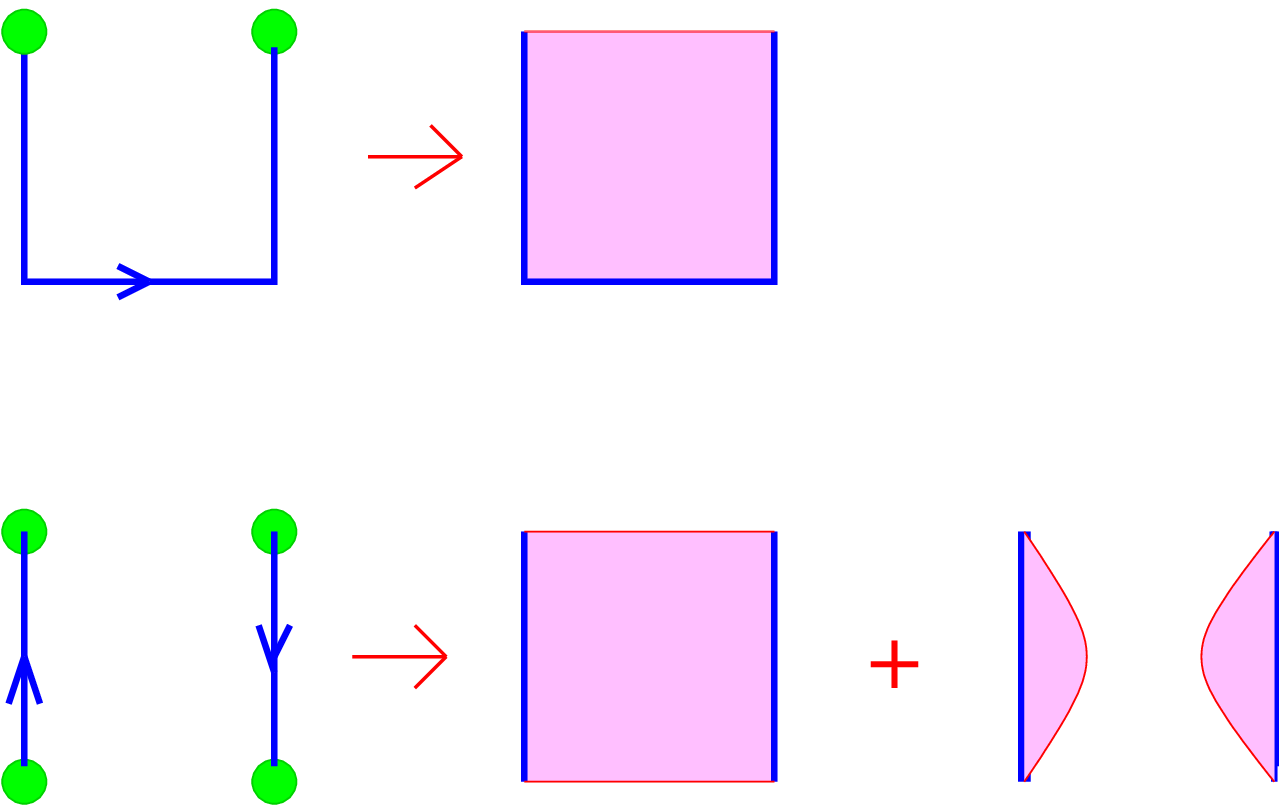}
%{}&
%\epsfxsize=.37\linewidth\epsfbox{5-dual.eps}
\\
%{}~&~
%\parbox[t]{.47\linewidth}{\small 1b) $dual ~polytope$.}
\end{array}\]
\end{figure}}

\begin{center}
\begin{minipage}{4.0truein}
               {  \footnotesize
                 \parindent=0pt
 Figure 4. Map to string world sheet of operators used
to determine the static potential in the SU(2) Higgs model.
The filled circles correspond to Higgs sources,
the thick lines represent Wilson lines or fixed boundaries,
the thin lines free boundaries.   }
                 \par
                 \end{minipage}\end{center}

\section{ String formulae}
\label{string}
The main role of the string picture of confinement is  to
fix the functional form of gauge operators in the infrared
limit.
In particular it is known \cite{aop} that, according to the
bosonic string model, a rectangular $R\times L$ Wilson loop
should behave asymptotically as the partition function of
$d-2$ free two-dimensional bosonic fields describing the transverse
oscillations of the world sheet with fixed b. c.:
\begin{equation}
\langle W(R,L)\rangle \propto e^{-\sigma R L-p(R+L)}
\left[\frac{\sqrt{R}}{\eta(\tau)} \right]^{\frac{d-2}{2}}
~,~~\tau=i \frac{L}{R}~,
\label{wilson}
\end{equation}
where $d$ is the spacetime dimension, $\sigma$ the
string tension, $p$ the perimeter term  and $\eta$ is the Dedekind function
\begin{equation}
\eta(\tau)=q^{1/24}\prod_{n=1}^\infty(1-q^n)~,~~
q\equiv \exp(2\pi i\tau)~.
\label{eta}
\end{equation}
This in turn gives the static potential
\begin{equation}
V(R)=-\lim_{L\to\infty}\frac{1}{L}\log\langle W(R,L)\rangle\label{defpot}
=\sigma R -\frac{(d-2)\pi}{24 R}+\dots
\label{luscher}
\end{equation}
where  dots indicate subleading terms.
These predictions have been tested to high accuracy for the $Z_2$ gauge
model in $d=3$ in \cite{z2}. Good agreement with the string model predictions
was also found for $SU(N)$ gauge theories in $d=3$ \cite{teper} and for
$SU(3)$ in $d=4$ \cite{beinlich}.
\par
Similar considerations apply to the confining phase of
finite temperature gauge theories. Here the relevant
observable is the correlator of Polyakov loops. Since the
lattice is finite and periodic in the time
direction, the Polyakov loop correlation will be described
by the
partition function of $d-2$ free bosons on a cylinder
with periodic b.c.
in the time direction and fixed b.c. on the two loops.
The partition function is therefore:
\begin{equation}
\langle P(0) P^{+}(R) \rangle \propto
Z_{DD}(R, L_t)
\propto\frac{e^{-\sigma_0 R L_t}}{\eta(\tau)^{d-2}}\label{zdd}
\end{equation}
where $\sigma_0$ is the
{\em zero temperature} string tension at the same coupling,
$L_t$ is the lattice extension in the time direction, that is the
inverse temperature, and now $\tau=i \frac{L_t}{2R}$.
Sec. \ref{quen} will be devoted to a comparison of Eq.~(\ref{zdd}) with lattice
quenched QCD data at finite temperature.
\par
We have seen in  Section \ref{topo} that in presence of dynamical
matter the Polyakov correlator contains an additional term
$[Z_{DN}(R_o,L_t)]^2$, where $Z_{DN}$ is the  partition
function on a cylinder with fixed b.c. on
the Polyakov loop side and free b.c. on the other
side, corresponding to a dynamically generated loop. The functional
form of $Z_{DN}$ is similar to that of Eq.~(\ref{zdd}), but now the
the $\eta$ function is  replaced by the  function
\begin{equation}
\tilde\eta(\tau_o)=q^{-1/48}\prod_{n=1}^\infty(1-
q^{n-\frac12})~,~~
\tau_o=\frac{iL_t}{2R_o},~q\equiv \exp(2\pi i\tau_o).
\label{etadn}
\end{equation}
The form af the two functions $\eta$ and $\tilde\eta$ can be
simply understood as
the inverse of the partition function of the normal modes
of vibration of the string. When both ends are fixed the allowed
frequencies are $\omega_n=\frac{\pi}R n$ and these generate
the infinite product of Eq.~(\ref{eta}); if one end is free
we have instead $\tilde\omega_n=\frac{\pi}R (n-\frac 12)$ which
generate the infinite product of Eq.~(\ref{etadn}).
The prefactor $q^{\frac 1{24}}$ in Eq.~(\ref{eta}) is directly
related to the zero point energy $E_o$ of the string
\begin{equation}
E_o=\sum_{n=1}^{~~~~\prime}\frac{\hbar\omega_n}2=
\frac{\hbar\pi}{2R}
\sum_n^{~~~~\prime}n=\frac{\hbar \pi}{2R} \zeta(-1)=
-\frac{\hbar\pi}{24R}~,
\end{equation}
where $\sum'$ denotes the $\zeta$-function regularized sum.
It is known that within such a regularization one can safely
handle the series as they were finite sums, then we can write
\cite{zaco}
\begin{equation}
\sum_{n=1}^{~~~~\prime}n=\sum_{n=1}^{~~~~\prime} 2n\,+\,
\sum_{n=1}^{~~~~\prime}(2n-1)
=2\zeta(-1)+2\sum_{n=1}^{~~~~\prime}(n-\frac 12)~~;
\end{equation}
hence
\begin{equation}
\sum_{n=1}^{~~~~\prime}\frac{\hb\tilde\omega_n}{2}=
\frac{\hbar\pi}{2R}\sum^{~~~~\prime}_{n=1}(n-\frac 12)=
\frac{\hbar\pi}{48R}~.
\end{equation}
This gives in turn the prefactor of $\tilde\eta$.
The identity $\tilde\eta(\tau)=\frac{\eta(\tau)}
{\eta(\tau/2)}$  allows us to write
\begin{equation}
Z_{DN}(R_o,L_t)=e^{-\sigma_0 R_o L_t}\left[\frac{\eta(\tau_o)}{\eta(\tau_o/2)}
\right]^{d-2} ~.
\label{zdn}
\end{equation}
Thus, assuming that the world sheet is in the normal phase,
we would expect that the Polyakov loop correlation function
in presence of dynamical quarks has the following effective
string description:
\begin{eqnarray}
\langle P(0) P^{+}(R) \rangle&&=c_1 Z_{DD}(R_o,L_t)+c_2
\left[Z_{DN}(R_o,L_t)\right]^2\nonumber\\
&=&c_1\frac{e^{-\sigma_0 R L_t}}{\eta(\tau)^{d-2}}+
\left(N_cN_f\right)^2
c_2 e^{-2\sigma_0 R_o L_t}\left[\frac{\eta(\tau_o)}{\eta(\tau_o/2)}
\right]^{2(d-2)}\label{model}
\end{eqnarray}
where $N_c$ and $N_f$ are respectively the number of colors and flavors of
light quarks.
The constants $c_1$ and $c_2$ and the length scale $R_o$ are not
predicted by the model and will have to be determined numerically. Notice
that the second term in Eq.~(\ref{model}) is just a constant at any fixed
temperature: therefore the actual predictions of the model concern the
temperature dependence of the potential. In Sec. \ref{unqu}
we will compare these predictions with lattice QCD data.
\vskip1.cm

\section{The static potential in quenched QCD at
finite temperature}
\label{quen}
If no dynamical quarks are present, the effective string prediction for
the Polyakov loop correlation is given by Eq.~(\ref{zdd}). For the
static potential we have in $d=4$
\begin{equation}
V(R)=-\frac{1}{L_t}\log \langle P(0) P^{+}(R) \rangle=\sigma_0 R+\frac{2}{L_t}
\log\eta\left(\frac{iL_t}{2R}\right)\label{potimp}
\end{equation}
We will be interested in the $2 R>L_t$ region, in which Eq.~(\ref{potimp}) is
conveniently rewritten in the equivalent form
\begin{equation}
V(R)=\sigma_0 R-\frac{\pi R}{3 L_t^2}+\frac{1}{L_t}\log\frac{2 R}{L_t}+
\frac{2}{L_t}
\sum_{n=1}^{\infty}\log\left(1-e^{-4\pi nR/L_t}\right)\label{potexp}
\end{equation}
Therefore at fixed temperature $1/L_t$ and asymptotically for large $R$
we have a temperature dependent string tension
\begin{equation}
\sigma(L_t)=\sigma_0-\frac{\pi}{3 L_t^2}
\label{sigmat}
\end{equation}
and a logarithmic term, plus exponentially suppressed subleading terms.
It is perhaps worth stressing that there is no $1/R$ term in this regime
\footnote{The L\"uscher term $\pi/12 R$ appears in the opposite
limit $2R\ll L_t$, as can be seen by rewriting Eq.~(\ref{potimp}) in
the other equivalent form
$$V(R)=\sigma_0 R-\frac{\pi}{12 R}+\frac{2}{L_t}
\sum_{n=1}^{\infty}\log\left(1-e^{-\pi nL_t/R}\right)
$$}.
\par
We want to compare the prediction Eq.~(\ref{potexp}) with Monte Carlo data
for quenched QCD at finite temperature. First let us stress some
important points:
\begin{enumerate}
\item{The free string picture we are using is an effective {\em infrared}
description: we expect it to describe the long distance behavior of the
potential (for a detailed discussion of this point in the zero temperature
case see Ref.~\cite{z2}). Therefore we expect to find a minimum distance
$R_{\rm string}$ above which Eq.~(\ref{potexp}) describes the data accurately.
Obviously there will be a distance $R_{\rm linear}$ above which the data
are well described also by the simple linear behavior  $V\sim \sigma R$.
The string model will be confirmed if $R_{\rm string}$ is sensibly
smaller than $R_{\rm linear}$.}
\item{The string model gives the finite temperature potential in terms
of the zero temperature string tension $\sigma_0$. Therefore the result
of the fit of the finite temperature potential should be compared to
the zero temperature string tension to confirm the string picture.}
\item{The free string picture must break down at temperatures near
the deconfinement point where string interactions are believed to
become important (see {\em e.g.} Ref.~\cite{olesen}). Therefore we need to
use lattice data at temperatures not too close to the critical one.}
\item{The main effect of string interactions is to renormalize the value
of the string tension $\sigma_0$ while leaving the functional form
Eq.~(\ref{potexp}) approximately unchanged. Therefore we expect string
interactions to introduce a systematic error in the evaluation of the zero
temperature string tension from finite $T$ data.}
\end{enumerate}
\par
Very precise data were obtained by the authors of Ref.~\cite{detar} at three
temperatures, with $T/T_c\sim 0.8,\  0.88\  {\rm and}\  0.94$. The first
of these data samples is ideal for our purpose.
We fitted those data to Eq.~(\ref{potimp}) and, for comparison,
to a simple linear behavior $V\sim \sigma R$.
Notice that these are both two parameter fits: the string contribution
to the potential does not contain any adjustable parameters.
Let us define  $R_{\rm string}$
as the value of $R$ above which the reduced $\chi^2$ of the fit is
$<1$, and $R_{\rm linear}$  as the corresponding distance for the linear
fit. The data are taken at $\beta=3.95$ and $L_t=4$. We obtain, in
terms of the lattice spacing $a$,
\begin{equation}
R_{\rm string}=3.3 a
\end{equation}
with $\chi^2=0.74$, and
\begin{equation}
R_{\rm linear}=4.5 a
\end{equation}
If we tried to fit the data for $R>R_{\rm string}$ to $V=\sigma R$ we
would obtain $\chi^2=8.6$.
\par
The fit with the string model potential and $R>R_{\rm string}$ gives
\begin{equation}
\sigma_0=0.22 a^{-2}\label{sigma0}
\end{equation}
This should be compared with the string tension at zero temperature at the
same value of the coupling $\beta=3.95$. Extrapolating the data of
Ref.~\cite{beinlich} to our $\beta$ we find $\sigma_0=0.24(1) a^{-2}$:
we conclude that the systematic error discussed above is still rather small
at $T/T_c=0.8$ (the statistical error on $\sigma_0$ given by our fit
is two order of magnitude smaller and hence not very meaningful).
\par
It is interesting to express $R_{\rm string}$ in physical units: using
\begin{equation}
\sigma_0=(420\ {\rm MeV})^2=4.41\  {\rm fm}^{-2}
\label{mev}
\end{equation}
we obtain
\begin{equation}
R_{\rm string}\sim 0.75\  {\rm fm}
\end{equation}
while using the string tension as the length scale
\begin{equation}
\sigma_0 R_{\rm string}^2\sim 2.4
\end{equation}
\par
If one studies the data samples closer to the critical temperature, at
$T/T_c\sim 0.88$ and $0.94$, one clearly sees the effect of string interactions
discussed above: while the
fit of the potential with the free string model potential is always very
good, the systematic error in the determination of $\sigma_0$ increases:
At $T/T_c\sim 0.88$ the fit to Eq.~(\ref{potexp})
gives  $\sigma_0\sim 0.18 a^{-2}$
when the Wilson loop value is \cite{beinlich} $0.197(8) a^{-2}$,
and for  $T/T_c\sim 0.94$  we obtain $\sigma_0\sim 0.13 a^{-2}$ instead
of $0.173(5) a^{-2}$.
\section{String breaking}
\label{unqu}
In this section we compare the prediction Eq.~(\ref{model}) to lattice data
taken by the authors of Ref.~\cite{detar} in finite temperature QCD with
two flavors of staggered dynamical quarks.
For the static potential Eq.~(\ref{model}) predicts, for $d=4$,
\begin{eqnarray}
V(R)&=&-\frac{1}{L_t}\log \langle P(0) P^{+}(R) \rangle=\nonumber\\
&-&\frac{1}{L_t}\log\left\{c_1\frac{e^{-\sigma_0 R L_t}}{\eta(\tau)^2}+
\left(N_cN_f\right)^2
c_2 e^{-2\sigma_0 R_o L_t}\left[\frac{\eta(\tau_o)}{\eta(\tau_o/2)}
\right]^4\right\}\nonumber\\
&=&-\log\left[\frac{e^{-\sigma_0 R L_t}}{\eta(\tau)^2}+
\left(N_cN_f\right)^2 c(R_o,L_t)\right]+A
\label{modpot}
\end{eqnarray}
where $A=-\frac{1}{L_t}\log c_1$ and
\begin{equation}
c(R_o,L_t)=\frac{c_2}{c_1} e^{-2\sigma_0 R_o L_t}\left[\frac{\eta(\tau_0)}
{\eta(\tau_0/2)}\right]^4\label{ratio}
\end{equation}
\par
The predictive content of the model lies in the dependence of $c$ on $L_t$.
To verify this prediction we fit the Monte Carlo data
for the static potential to the expression
\begin{equation}
V(R)=-\log\left[\frac{e^{-\sigma_0 R L_t}}{\eta(\tau)^2}+\left(N_cN_f
\right)^2 c\right]+A
\end{equation}
and verify that the dependence of $c$ on $L_t$ is actually described by
Eq.~(\ref{ratio}).
We used data taken by
the authors of Ref.~\cite{detar} at $L_t=4$ and four values of $\beta$
ranging from $\beta=5.1$ to $\beta=5.28$. The results
of the fits of the four data samples to Eq.~(\ref{ratio}) are reported in
Tab. 1, in lattice units:
\begin{table}[h]
\begin{center}
\begin{tabular}{|c|c|c|c|}
\hline
$\beta$&$\sigma_0 a^2$&$\left(N_cN_f\right)^2c$&$\chi^2$\\ \hline
5.10&0.594(80)&0.0151(68)&1.58\\ \hline
5.20&0.475(27)&0.0297(47)&1.11\\ \hline
5.25&0.407(19)&0.0653(72)&1.02\\ \hline
5.28&0.308(21)&0.189(25) &0.21\\ \hline
\end{tabular}
\caption{\em Fit of the lattice static potential to Eq.~(\ref{modpot}) for the
four values of $\beta$ considered}
\end{center}
\end{table}
\par
To verify Eq.~(\ref{ratio}) we need to trade the $\beta$ dependence at fixed
$L_t$ for a temperature dependence, by determining the $\beta$--dependent
lattice spacing $a(\beta)$.
We chose to use the values of $\sigma_0$ derived from our fit
to evaluate $a(\beta)$ (remember that $\sigma_0$ in Eq.~(\ref{model}) is
the zero temperature string tension). Therefore fixing $\sigma_0$ using
Eq.~(\ref{mev})
we obtain the lattice spacing at the four $\beta$ values and therefore the
inverse temperature in fm$^{-1}$. Using \cite{bernard} $T_c/\sqrt{\sigma_0}=
0.436(8)$ we can then determine the ratio $T/T_c$ for each $\beta$. These
values are reported in Tab. 2.
\begin{table}[h]
\begin{center}
\begin{tabular}{|c|c|c|c|}
\hline
$\beta$&$L_t\ ({\rm fm})$&$T/T_c$\\ \hline
5.10&1.468(96)&0.744(52)\\ \hline
5.20&1.313(28)&0.832(28)\\ \hline
5.25&1.215(28)&0.899(27)\\ \hline
5.28&1.057(36)&1.033(39)\\ \hline
\end{tabular}
\caption{\em Inverse temperature in fm and the ratio $T/T_c$
as computed from the
values of $\sigma_0$ for the four beta values.}
\end{center}
\end{table}
With the exception of the highest temperature, for which
we cannot trust the free string picture for the reasons mentioned in the
previous section,
the values of $T/T_c$ are
in good agreement with the ones quoted in Ref.~\cite{detar}.
\par
Now we can compare our results with Eq.~(\ref{ratio}). The latter contains
two free parameters, namely the length scale $R_o$ and the ratio $c_2/c_1$.
It is clearly safer to disregard the highest temperature point;
unfortunately in this way we have to perform a two parameter fit of three
data. Therefore the best we can hope at present is to show that the
Monte Carlo data are compatible with our model, while more data at
temperatures not too close to $T_c$ would be needed for a more conclusive test
of the model
\par
The fit of the three values of $c$ with Eq.~(\ref{ratio}) gives a
satisfactory reduced $\chi^2$ of 0.64. The length scale $R_o$ is
\begin{equation}
R_o=0.71(16)\  {\rm fm}
\end{equation}
The fact that this length scale turns out of the same order of magnitude
as the typical string breaking scale is of course encouraging. The constant
$\frac{c_2}{c_1}$ is affected by an error which is bigger than the value
of the constant itself: $\frac{c_2}{c_1}=4.2(6.4)$.
\par
If we push the model beyond its natural limits and use the parameters
given by our fit to predict the value of $c$ at the highest temperature, we
obtain for $\beta=5.28$ the value $\left(N_fN_c\right)^2c=0.20$, 
again in good agreement with
the numerical result. Even if this last result must be taken with all the
necessary caution, it is nevertheless a significant hint that the model is
realistic.
\par
We can conclude that the free string model of string breaking Eq.~(\ref{model})
is compatible with lattice result for finite temperature QCD, and certainly
reproduces at least the qualitative features of the phenomenon. More data
at moderate temperatures are certainly needed to draw a definite conclusion
about the correctness of the model.
\section{Concluding remarks}
\label{conc}
In this paper we conjectured that
the world sheet swept by the confining string  in presence of
matter fields belongs to the {\em normal} phase, characterized
by a large number of microscopic holes produced by the matter
fields, whose net effect is just to renormalize the string tension. 
This must be contrasted with the {\em tearing} phase, in which large holes 
dominate and drive the string tension between static sources to zero.
\par
This conjecture has two main consequences, both amenable to numerical
verification:
\begin{itemize}
\item 
String breaking cannot be detected by studying the large distance behavior 
of Wilson loops: even in presence of dynamical matter fields the Wilson
loop expectation value should behave at large distance like in the quenched
case, that is according to Eq.~(\ref{wilson}).   
\item
On the contrary, when one considers operators made of disjoint static sources
a simple topological argument shows that even if the world sheet is in the 
normal phase string breaking can be easily detected.
\end{itemize}
Both of these are in agreement with most of the 
rather large body of lattice data on the subject that has been 
published recently~\cite{data}-\cite{detar}. The only exception is
(2+1)-dimensional $SU(2)$ theory with two flavors of staggered 
fermions, where string breaking effects in the Wilson loop have been
observed~\cite{tr}. We suggest that
the mass of the quarks used in Ref.~\cite{tr} is large enough to make the world
sheet belong to the tearing phase. Indeed, the very same theory does not
show any sign of string breaking in the adjoint Wilson loop~\cite{Sa,PWa}
where the ``matter'' is massless by definition.   
\par
Based on this conjecture, we have proposed an effective string model of
string breaking for the simplest example of an operator constructed out of 
disjoint sources: the correlator of Polyakov loops in finite temperature
QCD. While the predictions of this model appear to be compatible with
the lattice data, more data would be needed for a conclusive 
assessment of its validity.
\par
One can envisage a number of numerical experiments to further check
the above conjecture and the effective string model.
Perhaps the simplest could be the analysis of the Polyakov
correlator in the adjoint representation at finite temperature.
In this case one can work directly in the pure Yang Mills
model and test accurately Eq.~(\ref{model}) as we did for
Eq.~(\ref{zdd}) in the quenched case.
\par
Also, it would be interesting to verify that the Wilson loop in 
presence of dynamical quarks actually behaves like in the quenched case,
that is according to Eq.~(\ref{wilson}). It is
however important to note that fuzzed Wilson loops should
be avoided in this kind of check, because the string quantum
fluctuations produce  a strong shape dependence which is
out of control in fuzzed loops. The $\eta$
function of Eq.~(\ref{wilson}) accounts for these fluctuations
only in the case of a rectangular shape. Likewise, analyses
based only on the asymptotic form  of the potential
given in Eq.~(\ref{luscher}) should be taken with caution, because
the logarithmic correction generated by the $\sqrt R$ term
is of the same order of magnitude as the L\"uscher term within 
the lattice size of current  simulations.
\vskip1.cm
\underline {Acknowledgements}
We would like to thank the authors of Ref.~\cite{detar} for kindly providing
us with their data, and M. Caselle, M. Hasenbusch, A. Lerda
and S. Vinti for useful discussions.

\end{document}